\documentclass[aps,prl,reprint,superscriptaddress]{revtex4-2}

\usepackage[T1]{fontenc}
\usepackage[utf8]{inputenc}
\usepackage{amsmath,amssymb,bm}
\usepackage{graphicx}
\usepackage{xcolor}

\definecolor{darkblue}{rgb}{0,0,0.6}
\definecolor{darkred}{rgb}{0.6,0,0}
\usepackage[colorlinks=true,urlcolor=darkblue, citecolor=darkblue, linkcolor=darkred, hyperfootnotes=false]{hyperref}

\newcommand{\dd}{\mathrm{d}}

\newcommand{\tD}{\widetilde D}
\newcommand{\tsigma}{\widetilde\sigma}
\newcommand{\tmu}{\widetilde\mu}
\newcommand{\trho}{\widetilde\rho}
\newcommand{\tQ}{\widetilde Q}
\newcommand{\jump}[1]{\left[#1\right]_{0^-}^{0^+}}
\newcommand{\Heav}{\Theta}

\newcommand{\moy}[1]{\left\langle #1 \right\rangle}
\def\rd{\tilde{\rho}}
\def\jd{\tilde{j}}
\def\rhotr{\Phi}

\begin{document}

\title{Exact Nonlinear Active Microrheology in Diffusive Single-File Systems}

\author{Aurélien Grabsch}
\affiliation{Sorbonne Universit\'e, CNRS, Laboratoire de Physique Th\'eorique de la Mati\`ere Condens\'ee (LPTMC), 4 Place Jussieu, 75005 Paris, France}

\author{Olivier Bénichou}
\affiliation{Sorbonne Universit\'e, CNRS, Laboratoire de Physique Th\'eorique de la Mati\`ere Condens\'ee (LPTMC), 4 Place Jussieu, 75005 Paris, France}

\date{\today}

\begin{abstract}
Active microrheology probes a crowded medium by forcing a tracer and measuring its response. In single-file transport, where particles cannot overtake, a constant force $F$ produces a subballistic displacement \(\langle X_t\rangle\simeq \sqrt{t}\,\xi(F)\) together with a persistent bath deformation. Despite decades of work, the exact nonlinear response at arbitrary force has remained confined to a few special solvable models. Here we remove this restriction by combining recent advances in hydrodynamic transport coefficients, a pressure-balance formulation of the local drive, and single-file duality, which eliminates the resulting moving boundary. This yields a closed exact boundary-value problem for general diffusive single files, determining both \(\xi(F)\) and the full density profile. For overdamped Brownian particles with general interactions, all microscopic interaction details enter only through the equilibrium equation of state, bringing realistic interacting systems, including finite-width quasi-one-dimensional channels, within exact reach. The solution also reveals universal global laws; in particular, the bath-density dipole is fixed by the applied force independently of the interaction potential.
\end{abstract}

\maketitle

\noindent\textit{Introduction.---}
Active microrheology probes a crowded medium by applying a controlled force to
a tracer and measuring the response of the probe~\cite{Bausch:1998,Habdas:2004,Meyer:2006,Squires:2005,Winter:2012,Winter:2013}. In ordinary fluids, this
response is often summarized by a force--velocity relation, which defines an
effective mobility.
Physically, the probe response is
coupled to  the deformation created by the driven tracer itself: particles
accumulate in front of it, a depleted region forms behind it, and in turn this
asymmetric density profile controls the tracer dynamics.

This  coupling is especially marked in single-file transport, where particles cannot overtake. This  is a paradigmatic regime of confined many-body dynamics, observed in systems ranging from molecular transport in zeolites to colloidal particles in narrow channels~\cite{Harris:1965,Arratia:1983,Karger:1992,Wei:2000,Lin:2005}. In this setting, moving the tracer requires the collective rearrangement of an ever-growing portion of the system. Consequently, even a constant force cannot sustain a stationary tracer velocity. Instead, at long times, \cite{Burlatsky:1992,Burlatsky:1996,Landim:1998}
\begin{equation}
\label{eq:mean_displacement}
    \langle X_t\rangle 
    \underset{t \to \infty}{\simeq}
    \sqrt t\,\xi(F)\:.
\end{equation}
The nonlinear response to an
arbitrary driving force is encoded in the amplitude $\xi(F)$,
which plays the role of a force--velocity relation in single-file transport.

At arbitrary drive, exact results for the mean tracer response and the
associated density profile have remained restricted to a few special solvable models. The most prominent example is the symmetric exclusion process (SEP), a minimal single-file model in which hard-core particles diffuse on a one-dimensional lattice and one tagged particle is biased~\cite{Burlatsky:1996,Landim:1998}. Another example is the random average process (RAP), in which particles move by random fractions of the available interparticle gaps~\cite{Cividini:2016}. Subsequent works have extended these results along several directions~\cite{Benichou:1999,Taloni:2008,Lizana:2010,Leibovich:2013,Demery:2014,Benichou:2016,Leitmann:2013,Leitmann:2017,Benichou:2018,Miron:2021}, and in particular have addressed fluctuations in these models~\cite{Landim:2000,Illien:2018,Cividini:2016,Kundu:2016,Dandekar:2022,Grabsch:2023b}.
These exact solutions rely on model-specific structures and do not provide a route to single-file systems with general interactions or realistic confinement.

\begin{figure*}
\centering
\includegraphics[width=0.8\textwidth]{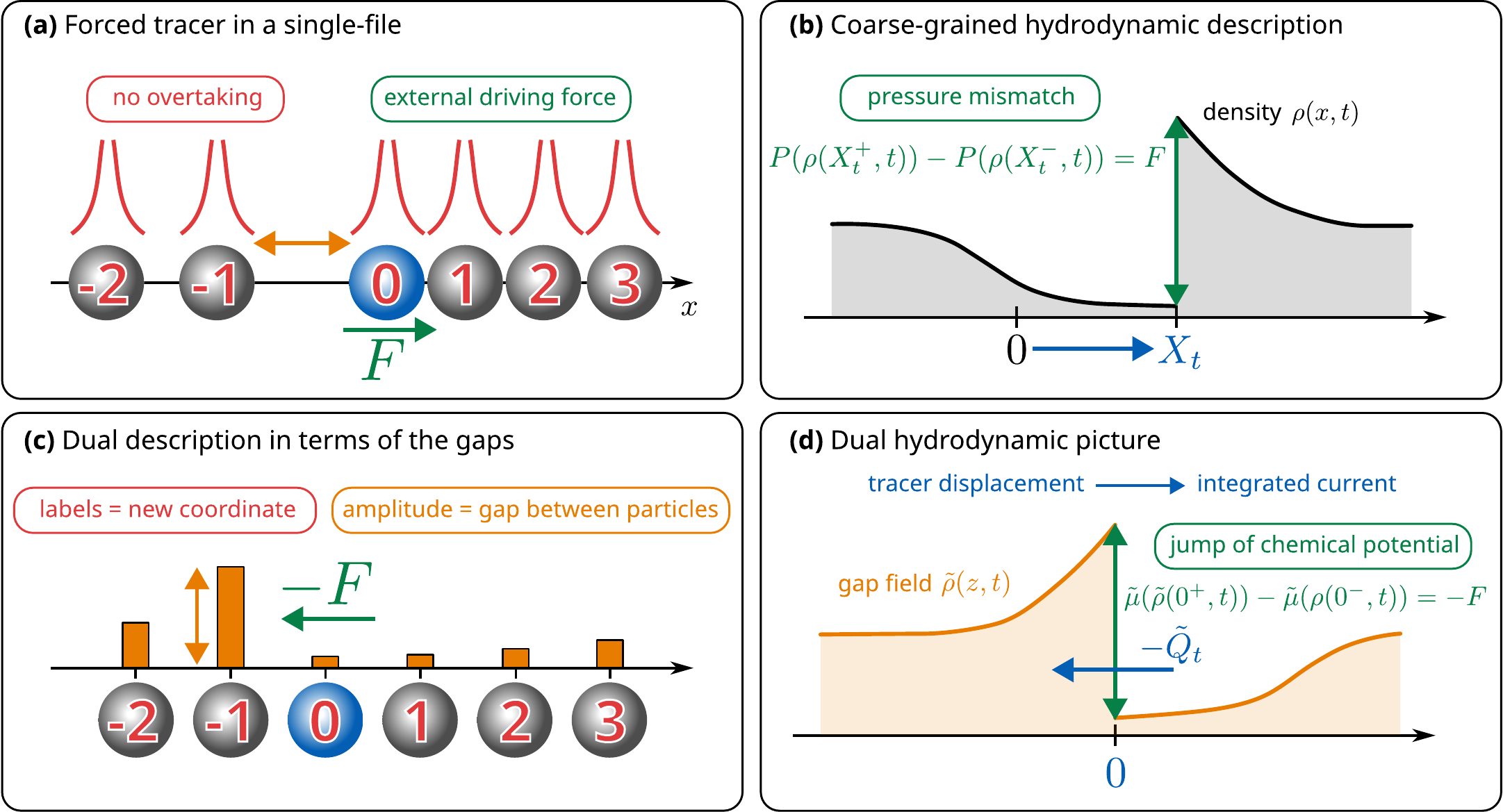}
\caption{\textbf{From nonlinear active microrheology to a fixed-boundary diffusion problem.} (a) A force $F$ drives a tracer in a single file.
(b) At the hydrodynamic scale, the force imposes a pressure mismatch across the
tracer at its unknown position $X_t$, defining a moving-boundary problem.
(c) The no-passing constraint allows the conserved particle label to be used as
a coordinate, fixing the tracer at $z=0$.
(d) The resulting dual hydrodynamic density $\tilde\rho$ obeys a diffusion problem on a fixed domain, with a chemical-potential jump at the origin, while the tracer displacement is
given by the integrated dual current.}
\label{fig:logic}
\end{figure*}

The question addressed here is whether this model-specific restriction can be overcome. Can one determine, in more realistic single-file systems, the tracer response and the induced density profile for an arbitrary driving force? We show that the answer is yes. This advance relies on assembling several ingredients that were only recently made available, and that remove the successive obstacles which had kept the arbitrary-drive problem restricted to a few special solvable models.~\cite{Burlatsky:1996,Landim:1998,Cividini:2016}.

\begin{figure*}
    \centering
    \includegraphics[width=\textwidth]{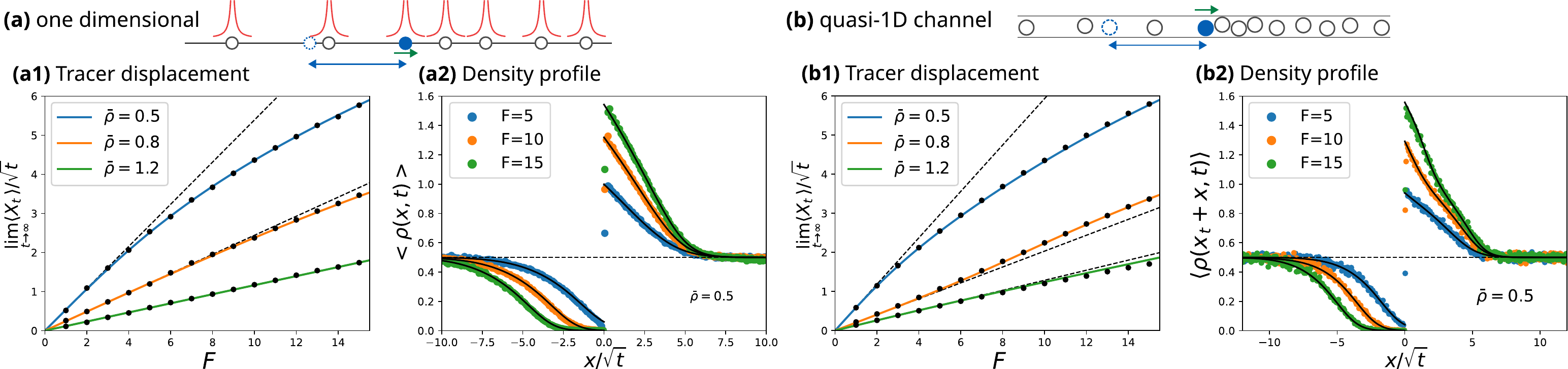}
    \caption{
    \textbf{Nonlinear active microrheology of interacting Brownian particles.}
(a) Strictly one-dimensional single-file system with pairwise Calogero
interactions $V(x)=1/x^2$.
(a1) Mean tracer-displacement amplitude $\xi(F)$ (see Eq. \eqref{eq:mean_displacement}) as a function of the driving
force $F$, for different mean densities $\bar\rho$. Solid lines are the
predictions of Eqs.~\eqref{eq:Q_ode}--\eqref{eq:xi_integral}, dashed lines are
the linear-response prediction~\eqref{eq:linear_response}, and symbols are
direct simulations of the Langevin dynamics~\eqref{eq:brownian_micro}, averaged
over $100$ realizations.
(a2) Mean density profile in the tracer frame at $\bar\rho=0.5$, for different
forces. Lines are the predictions of Eq.~\eqref{eq:profile_reconstruction} and
symbols are simulations averaged over $1000$ realizations.
(b) Finite-width single-file channel of Brownian particles interacting through
the Weeks--Chandler--Andersen potential~\cite{Weeks:1971}.
(b1) Mean tracer-displacement amplitude for different densities. Solid lines
are the theoretical predictions, dashed lines are the linear response, and
symbols are LAMMPS simulations~\cite{Thompson:2022} of $N=800$ particles in
channels of width $1.86$, with $\gamma=1/\mu_0=1$, $T=1$, and $m=0.1$,
averaged over $50$ realizations.
(b2) Mean density profile at $\bar\rho=0.5$ for different forces. Lines are the
theoretical predictions and symbols are simulations averaged over $1000$
realizations.}
    \label{fig:PlotsAll}
\end{figure*}

The first obstacle is the microscopic complexity of physical single-file systems with interactions and confinement. This complexity can be bypassed at long times, when the tracer motion involves the collective rearrangement of an ever-growing region of the file and naturally admits a hydrodynamic description. At the fluctuating hydrodynamic level, a diffusive single file is specified by its collective diffusivity \(D(\rho)\) and mobility \(\sigma(\rho)\)~\cite{Spohn:1991,Bertini:2015}. Recent results show how these coefficients can be determined for systems with general interactions~\cite{Grabsch:2025b,Grabsch:2026}, and for finite-width quasi-one-dimensional confinement~\cite{Benichou:2026}. This is the input that brings physical systems within reach, as illustrated below for strictly one-dimensional interacting Brownian particles and for finite-width channels in
Fig.~\ref{fig:PlotsAll}.
The second obstacle is that the driving force acts locally on a single tracer, whereas the hydrodynamic description concerns coarse-grained fields. 
This difficulty can nevertheless be overcome: at large scales, the local drive is encoded in a closed boundary condition, namely a pressure mismatch across the tracer.
The last obstacle is that this pressure condition is imposed at the unknown tracer position, thus defining a moving-boundary problem. We rely on the single-file duality transformation~\cite{Rizkallah:2022} to overcome this difficulty: the tracer becomes the fixed origin of the label coordinate, its displacement becomes a dual integrated current, and the pressure mismatch becomes a fixed chemical-potential jump for a nonlinear diffusion equation. The combination of these ingredients, summarized in Fig.~\ref{fig:logic}, yields a closed boundary-value problem whose solution gives both the exact force--velocity relation \(\xi(F)\) and the full density profile in the tracer frame.

\noindent\textit{Model and hydrodynamic description.---}
We consider an infinite diffusive single file in which particle order is preserved. Microscopic interactions and confinement are left unspecified at this stage ; in particular, the file may be strictly one-dimensional or realized in a narrow quasi-one-dimensional channel. The tagged particle is initially at the origin, its position at time $t$ is $X_t$, and it is the only particle subjected to an external force $F$. At large scales, this system is described by  coarse-grained line-density $\rho(x,t)$, which obeys the fluctuating hydrodynamic equation~\cite{Spohn:1991,Bertini:2015,Benichou:2026}
\begin{equation}
  \partial_t\rho+\partial_x j=0
  \:,
  \qquad
  j=-D(\rho)\partial_x\rho+
  \sqrt{\frac{\sigma(\rho)}{\Lambda}}\,\eta \:,
  \label{eq:fh_direct}
\end{equation}
where $D(\rho)$ is the collective diffusivity, $\sigma(\rho)$ is the mobility, $\eta$ is a Gaussian white noise, and $\Lambda\gg1$ is the hydrodynamic scale. The same functions $D$ and $\sigma$ determine the equilibrium pressure through  \cite{Hill:1986,Derrida:2025a}
\begin{equation}
  P(\rho)= 2 k_{\mathrm{B}} T \int^\rho \frac{rD(r)}{\sigma(r)}\,\dd r
  \:.
  \label{eq:pressure_transport}
\end{equation}
We allow a step preparation, with densities $\rho_+$ and $\rho_-$ far to the right and left of the tracer. The standard active-microrheology setting is recovered for $\rho_+=\rho_- \equiv \bar\rho$.

\noindent\textit{Driven tracer at the hydrodynamic level.---}
The fluctuating hydrodynamic equations~\eqref{eq:fh_direct} describe the global dynamics of all the particles. For an undriven tracer, its displacement can be recovered from these fields using the Eulerian--Lagrangian correspondence~\cite{Lamb:1932},
\begin{equation}
\label{eq:EulerPos}
    \frac{\dd X_t}{\dd t}
    = v(X_t,t)
    = \frac{j(X_t,t)}{\rho(X_t,t)}
    \:.
\end{equation}
Due to the conservation of the particle's order, this relation also holds for a driven tracer at the hydrodynamic scale, since it has the same displacement as all the particles in the same mesoscopic fluid cell.
It remains to determine how the driving force $F$ enters this hydrodynamic description.

This is in principle a difficult task: at the microscopic level the force modifies the local dynamics around the tracer, resulting in a complex many-body problem for which no general exact solution is available. Nevertheless, in the case of the SEP, closed matching conditions have been derived for the hydrodynamic field $\rho(x,t)$ at the position of the tracer~\cite{Burlatsky:1996,Landim:1998,Dandekar:2022,Grabsch:2023b}, but not in a physically transparent form. 

The hydrodynamic approach actually provides a physical derivation of a general matching condition. Indeed, each mesoscopic fluid region can be treated as locally at equilibrium, with density $\rho(x,t)$. The regions immediately to the right and left of the tracer therefore exert forces determined by the one-dimensional thermodynamic pressures $P(\rho(X_t^+,t))$ and $P(\rho(X_t^-,t))$. In addition, since the tracer velocity vanishes on hydrodynamic time scales,  $\dd\langle X_t\rangle/\dd t\sim t^{-1/2} \to 0$, the tracer becomes quasi-static at long times. The force balance then yields
\begin{equation}
  P\!\left(\rho(X_t^+,t)\right)
  -P\!\left(\rho(X_t^-,t)\right)=F .
  \label{eq:pressure_jump}
\end{equation}
Equation~\eqref{eq:pressure_jump} was already used in the related problem of force-induced unbinding in single files~\cite{Poncet:2018b}, where it was motivated from the SEP matching condition, and further discussed in Ref.~\cite{Poncet:2020}. Here, we derive it directly from the hydrodynamic force balance and use it as the general matching condition for driven diffusive single files.

The remaining difficulty is that the matching condition~\eqref{eq:pressure_jump} is imposed at the tracer position $X_t$, which is itself determined self-consistently by~\eqref{eq:EulerPos}, thus defining a moving-boundary problem. We now remove this difficulty by relying on the single-file duality.

\noindent\textit{Single-file duality.---}
Thanks to the no-passing constraint, the conserved particle label can be used as a coordinate measured from the tracer. Choosing the tracer label to be $z=0$ thus makes it the fixed origin of this new coordinate. Remarkably, a dual hydrodynamic description can be constructed based on this new coordinate. If $x(z,t)$ is the physical position of particle with label $z$ at time $t$ we define the dual density
\begin{equation}
  \rd(z,t)=\frac{\partial x}{\partial z}=\frac{1}{\rho(x(z,t),t)} 
  \:,
  \label{eq:dual_density}
\end{equation}
and the associated current $\jd(z,t)$. The new field $\rd$ thus represents the local interparticle spacing, while $\jd(z,t)$ is the associated spacing current.
The position of the particles can thus be determined as
\begin{equation}
    \label{eq:DefPos}
    x(z,t) = \int_0^z \rd(z',t) \dd z'
    - \int_0^t \jd(0,t') \dd t'
    \:,
\end{equation}
where the first term gives the distance between particle $z$ and the tracer, while the second term gives the tracer position. The new fields obey the same fluctuating hydrodynamic equation~\eqref{eq:fh_direct}, but with dual transport coefficients given by~\cite{Rizkallah:2022}
\begin{equation}
  \tD(\rd)=\frac{1}{\rd^2}
  D\!\left(\frac{1}{\rd}\right),
  \qquad
  \tsigma(\rd)= \rd \,\sigma\!\left(\frac{1}{\rd}\right)
  \:,
  \label{eq:dual_transport}
\end{equation}
and the dual chemical potential is
\begin{equation}
  \tmu(\rd)=
  2 k_{\mathrm{B}}T
  \int^{\rd} \frac{\tD(r)}{\tsigma(r)}\,\dd r 
  \:.
  \label{eq:dual_mu}
\end{equation}
Equations~\eqref{eq:pressure_transport} and~\eqref{eq:dual_mu} imply $P(1/\rd)=-\tmu(\rd)+{\rm const.}$, hence the pressure condition becomes a fixed jump condition at the origin,
\begin{equation}
  \tmu(\rd(0^+,t))-\tmu(\rd(0^-,t))=-F .
  \label{eq:dual_jump}
\end{equation}
The tracer displacement is also fixed by the dual fields via Eq.~\eqref{eq:DefPos} at $z=0$, which expresses it as the opposite of the integrated dual current. The continuity relation $\partial_t \rd + \partial_z \jd = 0$ allows us to express it in terms of $\rd$ only,
\begin{equation}
  X_t=-\tQ_t,
  \qquad
  \tQ_t=\int_0^\infty
  \left[\rd(z,t)-\rd(z,0)\right]\dd z 
  \:.
  \label{eq:current_identity}
\end{equation}
Thus the driven-tracer problem is converted into a diffusion problem on two fixed half-lines, coupled by the jump condition~\eqref{eq:dual_jump} and by current conservation at the origin (see Supplementary Material (SM)~\cite{SM} for a detailed justification of this point).

\noindent\textit{Exact nonlinear response.---}
Let $\trho_\pm=1/\rho_\pm$ the dual asymptotic densities. Since the noise term in the fluctuating hydrodynamic equation~\eqref{eq:fh_direct} is small, the mean dual density $q(z,t) = \moy{\rd(z,t)}$ solves
\begin{align}
  \partial_t q
  &=\partial_z\!\left[\tD(q)\partial_z q\right]
  \:,
  \qquad z \neq 0
  \:,
  \label{eq:mean_dual_bulk}\\
  q(z,0)&=\trho_+\Heav(z)+\trho_-\Heav(-z)
  \:,
  \label{eq:mean_dual_ic}\\
  \jump{\tmu(q)}&=-F
  \:,
  \qquad
  \jump{\tD(q)\partial_z q}=0
  \:.
  \label{eq:mean_dual_contact}
\end{align}
The solution of these equations is self-similar, with $q(z,t)=Q(u = z/\sqrt{t})$, and is determined by
\begin{align}
  0&=\frac{\dd}{\dd u}\left[\tD(Q)\frac{\dd Q}{\dd u}\right]
  +\frac {u}{2}\frac{\dd Q}{\dd u}
  \:,
  \label{eq:Q_ode}\\
  Q(+\infty)&=\trho_+
  \:,
  \qquad Q(-\infty)=\trho_-
  \:,
  \label{eq:Q_far}\\
  \jump{\tmu(Q)}&=-F
  \:,
  \qquad
  \jump{\tD(Q)Q'}=0 
  \: .
  \label{eq:Q_contact}
\end{align}
These equations fully determine the mean dual density profile $q(z,t)$, from which the mean displacement of the tracer follows from~\eqref{eq:current_identity}. The square root dependence on time~\eqref{eq:mean_displacement} follows from the self-similar form, and the prefactor reads
\begin{equation}
  \xi(F)=-\int_0^\infty\left[Q(u)-\trho_+\right]\dd u
  = 2 \tD(Q) Q' \Big|_0
  \:,
  \label{eq:xi_integral}
\end{equation}
where we used~\eqref{eq:Q_ode} for the last equality.

Equations~\eqref{eq:Q_ode}--\eqref{eq:xi_integral} provide a closed force-response relation for any diffusive single-file system. In the specific case of the SEP, corresponding to $D(\rho)=1$ and $\sigma(\rho)=2\rho(1-\rho)$, we recover the equations derived in Ref.~\cite{Landim:1998}. For arbitrary $D$ and $\sigma$, these equations define a closed one-dimensional boundary-value problem that can be solved numerically at any force (see SM~\cite{SM}). They also yield explicit analytic expansions in limiting cases. For instance, for a weak force and a flat initial density $\rho_+=\rho_-=\bar\rho$, we obtain
\begin{equation}
  \xi(F)=
  \frac{\sigma(\bar\rho)}{\bar\rho^2\sqrt{4\pi D(\bar\rho)}} \frac{F}{k_{\mathrm{B}}T}
  +O(F^3)
  \:.
  \label{eq:linear_response}
\end{equation}
The linear term satisfies the equilibrium fluctuation--dissipation relation $\left.\partial_F\moy{X_t}\right|_{F=0}
=\moy{X_t^2}_{F=0}/(2k_{\mathrm B}T)$, using the exact equilibrium tracer variance~\cite{Kollmann:2003,Krapivsky:2014,Krapivsky:2015a}.
The expansion can be pushed further analytically to obtain the cubic correction explicitly, as shown in the SM~\cite{SM}.

\noindent\textit{Density profile.---}
Although the tracer response is the primary microrheological observable, the same solution gives the bath deformation that controls this response. The mean density in the reference frame of the tracer,
\begin{equation}
    \moy{\rho(X_t+x,t)} = 
    \rhotr \left(y = \frac{x}{\sqrt{t}} \right)
    \:,
\end{equation}
directly follows from the solution $Q$ of Eqs.~\eqref{eq:Q_ode}--\eqref{eq:Q_contact} by the duality mapping~\eqref{eq:dual_density},
\begin{equation}
  \rhotr(y(u))=\frac{1}{Q(u)} 
  \:,
  \quad
  y(u)=\int_0^u Q(u')\,\dd u'
  \:.
  \label{eq:profile_reconstruction}
\end{equation}
This is due to the fact that the noise is small in the fluctuating hydrodynamic equation~\eqref{eq:fh_direct}, therefore
$\moy{\rd(z,t)} = 1/\moy{\rho(\moy{x(z,t)},t)}$. The bath profile in the tracer frame thus is reconstructed from the same solution $Q(u)$ that determines the displacement of the tracer~\eqref{eq:xi_integral}.

\noindent\textit{Interacting Brownian particles.---}
We finally apply this general formalism to the important and experimentally relevant case of overdamped Brownian particles with general interactions for which the large-scale dynamics is diffusive~\footnote{This is the case for potentials that decay faster than $1/x$, for which the system has a standard thermodynamic (extensive energy).},
\begin{equation}
  \frac{\dd x_n}{\dd t}
  =-\mu_0\sum_{m\ne n}V'(x_n-x_m)+
  \mu_0
  F
  \delta_{n,0}
  +\sqrt{2\mu_0 k_BT}\,\eta_n,
  \label{eq:brownian_micro}
\end{equation}
where $\mu_0$ is the bare mobility and the $\eta_n$ are independent unit Gaussian white noises.
For such systems, all the microscopic details enter the transport coefficients only through the equation of state.
More precisely~\cite{Lekkerkerker:1981,Cichocki:1991,Butta:1999,Felderhof:2009,Grabsch:2025b,Grabsch:2026},
\begin{equation}
  D(\rho)=\mu_0 \,P'(\rho) \:,
  \qquad
  \sigma(\rho)=2\mu_0 k_{\mathrm{B}} T\rho \:,
  \label{eq:brownian_transport}
\end{equation}
where $P$ is the equilibrium pressure for a system at density $\rho$. This reduces the determination of the transport coefficients to the computation of the equilibrium pressure. This can be done explicitly for specific potentials, such as hard rods or Calogero $V(x) = a/x^2$~\cite{Calogero:1975,Moser:1975,Choquard:2000,Lewin:2022}, or approximately by relying on standard techniques such as the virial expansion~\cite{Hill:1986,Hansen:2005}.
In dual variables this gives
\begin{align}
  \tD(q)&=\mu_0 \,\tmu'(q)
  \:,
  &
  \tsigma(q)&=2\mu_0 k_{\mathrm{B}}T
  \:,
  \label{eq:brownian_dual_a}\\
  \tmu(q)&=-P(1/q)+{\rm const.}
  \label{eq:brownian_dual_b}
\end{align}
The dual mobility does not depend on the density and all the information on the interaction is encoded in the dual chemical potential $\tmu$.
Inserted into Eqs.~\eqref{eq:Q_ode}--\eqref{eq:Q_contact}, these relations fully determine the self-similar function $Q(u)$, from which the tracer displacement
and the density profile follow through Eqs.~\eqref{eq:mean_displacement},~\eqref{eq:xi_integral}, and~\eqref{eq:profile_reconstruction}.
This gives the nonlinear active-microrheology response for arbitrary Brownian interactions from the equilibrium equation of state alone. We stress that no closure approximation for the driven nonequilibrium bath profile is introduced so that these results are exact.
This is the practical gain of the hydrodynamic formulation: the driven many-body problem is reduced to the computation of the equilibrium pressure, which is then inserted into the universal nonlinear boundary-value problem~\eqref{eq:Q_ode}-\eqref{eq:Q_contact}.

We compare in 
Fig.~\ref{fig:PlotsAll}(a)
our predictions for the mean displacement~\eqref{eq:mean_displacement},\eqref{eq:xi_integral} and the density profile~\eqref{eq:profile_reconstruction} against numerical simulations of the Langevin equations~\eqref{eq:brownian_micro}. The plots show an excellent agreement for all densities and forces.

\noindent\textit{Finite-width single-file system.---}
Importantly, this formalism also describes the more realistic case of Brownian particles in a finite-width channel, chosen small enough to preserve the single-file constraint. The effective one-dimensional equations~\eqref{eq:Q_ode}-\eqref{eq:xi_integral} still hold, with the same transport coefficients~\eqref{eq:brownian_dual_a}-\eqref{eq:brownian_dual_b}~\cite{Benichou:2026}. The equation of state can be computed exactly for nearest-neighbor interactions using a transfer matrix formalism~\cite{Kofke:1993,Montero:2023,Franosch:2024}. The results, given in
Fig.~\ref{fig:PlotsAll}(b) 
show an excellent agreement with our predictions. The small discrepancy observed at high density is due to the nearest-neighbor approximation which becomes inaccurate at high density.

Qualitatively, the behaviors of the displacement $\xi(F)$ and the bath density profile are similar in the strictly-1D and quasi-1D situations: the forced displacement of the tracer causes an accumulation of particles in front of it, and a depletion behind. The amplitude of this imbalance increases with the driving force, and so does the mean displacement. This effect is also reduced when the density is increased, since the same displacement requires  pushing more particles.

Quantitatively, there are important differences to notice in the two cases. In the strictly-1D case 
(Fig.~\ref{fig:PlotsAll}(a)),
the displacement is always smaller than the linear response prediction~\eqref{eq:linear_response}. However, in the quasi-1D case 
(Fig.~\ref{fig:PlotsAll}(b)),
the displacement can be larger than the linear response for some densities. This can be traced to the nonmonotonic dependence of the diffusion coefficient
$D(\rho)$ on density in this case~\cite{Benichou:2026}. This also affects the
density profiles, which display a nonconvex behavior.

\noindent\textit{Universal laws of nonlinear active microrheology.---}
Despite these quantitative differences, the exact solution reveals two global identities valid for arbitrary $D$ and $\sigma$. For a homogeneous bath, $\rho_+=\rho_-=\bar\rho$,
\begin{equation}
  \int_0^\infty [\rhotr(y) - \bar\rho]\dd y
  =-\int_{-\infty}^{0}[\rhotr(y) - \bar\rho]\dd y=\bar\rho\,\xi(F),
  \label{eq:mass_sum_rule}
\end{equation}
\begin{equation}
    \int_{-\infty}^{\infty}
    y[\rhotr(y) - \bar\rho]\dd y
  =\int_{\rhotr(0^-)}^{\rhotr(0^+)}D(r)\,\dd r.
  \label{eq:general_dipole_sum_rule}
\end{equation}
The first identity is an exact mass balance: the tracer response equals the scaled excess mass accumulated ahead of the tracer, or equivalently the deficit left behind. The second relates the dipole moment of the same deformation to the collective diffusivity between the two contact densities. Together, these identities reveal exact global constraints on the bath deformation, beyond the system-dependent details of its spatial profile.

For Brownian dynamics, the diffusion coefficient~\eqref{eq:brownian_transport} and the pressure balance~\eqref{eq:pressure_jump} collapse Eq.~\eqref{eq:general_dipole_sum_rule} to
\begin{equation}
  \int_{-\infty}^{\infty}y
  [\rhotr(y) - \bar\rho]
  \dd y=\mu_0 F.
  \label{eq:brownian_dipole_sum_rule}
\end{equation}
Remarkably, the density dipole is therefore fixed by the applied force only, independently of the interaction potential. Interactions affect $\xi(F)$ only by redistributing this imposed dipole in space, thereby changing the displaced mass entering Eq.~\eqref{eq:mass_sum_rule}.
The bath deformation thus encodes both sides of active microrheology: its displaced mass measures the response, while its dipole measures the drive. For Brownian files this provides a direct interaction-independent test of the nonlinear theory, requiring no knowledge of the interaction potential.

\noindent\textit{Conclusion.---}
We have obtained the ``force-velocity'' relation $\xi(F)$ and the associated bath density profile for general single-file systems. Our analysis goes beyond previous results restricted to specific microscopic models like the SEP~\cite{Landim:1998} by relying on a large scale hydrodynamic description in terms of the transport coefficients $D(\rho)$ and $\sigma(\rho)$.
This description includes lattice gases like the SEP, but importantly more realistic models like interacting Brownian particles. The solution also reveals exact global constraints on the bath deformation;
for Brownian files, in particular, its density dipole is fixed by the applied
force independently of the interactions.

Our work opens the question of the characterization of the fluctuations around the mean displacement~\eqref{eq:mean_displacement}. Even for the SEP, the variance of a driven tracer is not known at arbitrary force; only its small-drive behavior has recently been determined~\cite{Grabsch:2023b}.
The formalism presented in this work in principle gives a way to access the full distribution of $X_t$ for general systems, as shown in SM~\cite{SM}. However, solving these equations, even for a simple model like the SEP, remains a challenging open problem.


%

\end{document}


\setcounter{equation}{0}
\setcounter{figure}{0}
\setcounter{table}{0}
\setcounter{page}{1}
\makeatletter
\renewcommand{\theequation}{S\arabic{equation}}
\renewcommand{\thefigure}{S\arabic{figure}}
\renewcommand{\bibnumfmt}[1]{[S#1]}
\renewcommand{\citenumfont}[1]{S#1}

\title{Supplemental Material for\texorpdfstring{\\}{} Exact Nonlinear Active Microrheology in Diffusive Single-File Systems}

\maketitle

\tableofcontents

\section{Fluctuating hydrodynamics with a forced tracer}

Let us denote $\{ x_n(t) \}$ the positions of the particles at time $t$. We denote $X_t = x_0(t)$ the position of the tracer at time $t$, with $X_0 = 0$. We introduce the microscopic density $\rhom$ and current $\jm$ of the other particles as
\begin{equation}
    \label{eq:DefMicroFieldsSM}
    \rhom(x,t) = \sum_{n \neq 0} \delta \big(
    x_n(t) - x
    \big)
    \:,
    \quad
    \jm(x,t) = \sum_{n \neq 0}
    \dt{x_n}{t}
    \delta \big(
    x_n(t) - x
    \big)
    \:,
\end{equation}
which satisfy the continuity relation $\partial_t \rhom + \partial_x \jm = 0$.
We define the macroscopic coarse-grained density $\rho(x,t)$ and current $j(x,t)$ as
\begin{equation}
    \rho(x,t)
    = \frac{1}{\ell} \int_0^\ell \rhom(\Lambda x + x', \Lambda^2 t)
    \dd x'
    \:,
    \quad
    j(x,t) =
    \frac{\Lambda}{\ell} \int_0^\ell 
    \jm(\Lambda x + x', \Lambda^2 t)
    \dd x'
    \:,
\end{equation}
where $\ell$ is a mesoscopic scale on which the system can be seen as locally at equilibrium. This will be important in the following. The rescaling parameter $\Lambda \gg 1$ is the ratio between the macroscopic and microscopic scales. Note that we have performed a diffusive rescaling of space and time.

The macroscopic fields obey the fluctuating hydrodynamics equations~\cite{Spohn:1991,Bertini:2015}
\begin{equation}
  \partial_t\rho+\partial_x j=0
  \:,
  \qquad
  j=-D(\rho)\partial_x\rho+
  \sqrt{\frac{\sigma(\rho)}{\Lambda}}\,\eta \:,
  \label{eq:fhSM}
\end{equation}
where $\eta$ is a Gaussian white noise in space and time,
\begin{equation}
    \moy{\eta(x,t) \eta(x',t')} =
    \delta(x-x') \delta(t-t')
    \:.
\end{equation}
All the microscopic details (dynamics, interaction, etc) are encoded in the two transport coefficients: the collective diffusion coefficient $D(\rho)$ and the mobility $\sigma(\rho)$.

\subsection{Matching conditions}

The macroscopic density $\rho(x,t)$ and current $j(x,t)$ encode the information on all the particles except the tracer. The displacement $X_t$ of the tracer can be determined from the macroscopic fields by using the correspondence between the Eulerian and Lagrangian pictures, which states that~\cite{Lamb:1932}
\begin{equation}
    \dt{X_t}{t} = v_0(X_t,t)
    \:,
\end{equation}
with $v_0$ the velocity field. Since $j_0 = \rho_0 v_0$, we can write
\begin{equation}
    \label{eq:DefDisplSM}
    \rhom(X_t,t) \dt{X_t}{t} = \jm(X_t,t)
    \quad \Rightarrow \quad
    \rho(Y_t,t) \dt{Y_t}{t}  = j(Y_t,t) 
    \:,
    \quad \text{where} \quad
    Y_t = \frac{X_{\Lambda^2 t}}{\Lambda}
    \:.
\end{equation}
These equations define the macroscopic position $Y_t$ of the tracer from the macroscopic fields, but do not enforce the action of the external force $F$ onto the tracer. For this we need enforce this driving force at the level of the macroscopic fields. This can be done by investigating the equation of motion of the tracer.
Since the displacement of the tracer behaves as $\moy{X_t} \sim \sqrt{t}$ for large $t$, we have that $\dt{\moy{X_t}}{t} \sim 1/\sqrt{t}$. Hence, $\dt{X_t}{t} \big|_{\Lambda^2 t} \to 0$ for $\Lambda \to \infty$. Therefore, at large scale, the tracer can be treated as having no velocity and thus being at equilibrium.
Additionally, every other part of the system can be treated to be locally at equilibrium. Thus, the force exerted by all the particles to the right of the tracer is $-P(\rho(Y_t^+,t))$, while the force exerted by the particles to the left if $P(\rho(Y_t^-),t)$. The force balance on the tracer thus implies a matching condition for the macroscopic fields
\begin{equation}
    \label{eq:MatchBiasSM}
    P\big(\rho(Y_t^+,t) \big) 
    - P\big(\rho(Y_t^-,t) \big) = F
    \:.
\end{equation}
The action of the external driving force can thus be implemented at the macroscopic level by a discontinuity in the pressure at the position of the tracer. Note that the pressure $P(\rho)$ is expressed in terms of the transport coefficients as
\begin{equation}
\label{eq:DefPSM}
    \partial_\rho P(\rho) = \frac{2 k_{\mathrm{B}} T \rho D(\rho)}{\sigma(\rho)}
    \:,
\end{equation}
since by denoting $f(\rho)$ the free energy density~\cite{Hill:1986SM}
\begin{equation}
    P(\rho) = 
   - \left( 
    \frac{\partial F}{\partial V}
    \right)_{T,N}
    = - \left( 
    \frac{\partial [V f(N/V)]}{\partial V}
    \right)_{T,N}
    = \rho f'(\rho) - f(\rho)
    \quad \Rightarrow \quad
    P'(\rho) = \rho f''(\rho)
    \:,
\end{equation}
and $f''(\rho) = 2 k_{\mathrm{B}} T D(\rho) / \sigma(\rho)$~\cite{Derrida:2025aSM}.

\medskip 

Furthermore, the definition of the displacement $Y_t$~\eqref{eq:DefDisplSM} can be written on each side of the tracer, which thus implies
\begin{equation}
    \label{eq:ContXtSM}
    \frac{j(Y_t^+,t)}{\rho(Y_t^+,t)}
    = \frac{j(Y_t^-,t)}{\rho(Y_t^-,t)}
    \:.
\end{equation}
Together, the evolution equations~\eqref{eq:fhSM} and the matching conditions~\eqref{eq:MatchBiasSM}-\eqref{eq:ContXtSM} fully determine the macroscopic dynamics of the system.

\medskip

Note that the definition~\eqref{eq:DefDisplSM} of the displacement of the tracer is equivalent to the one used recently in the context of macroscopic fluctuation theory~\cite{Krapivsky:2014,Krapivsky:2015a}
\begin{equation}
    \int_0^{Y_t} \rho(x,t) \dd x
    = \int_0^{\infty} \big[
    \rho(x,t) - \rho(x,0)
    \big] \dd x
    \:.
\end{equation}
Indeed, taking a derivative of this definition with respect to time yields
\begin{equation}
    \dt{Y_t}{t} \rho(Y_t,t)
    + \int_0^{Y_t} \partial_t \rho(x,t) \dd x
    = \int_0^\infty \partial_t \rho(x,t) \dd x
    \:.
\end{equation}
Using the continuity equation~\eqref{eq:fhSM}, we can perform the integration to get
\begin{equation}
    \dt{Y_t}{t} \rho(Y_t,t)
    - j(Y_t,t) = 0
    \:,
\end{equation}
which is exactly~\eqref{eq:DefDisplSM}.

\subsection{Dual description}

The study of the macroscopic fields is difficult, due to the presence of the matching conditions~\eqref{eq:MatchBiasSM}-\eqref{eq:ContXtSM} at the unknown position $Y_t$, determined from the fields $\rho$ and $j$ themselves via~\eqref{eq:DefDisplSM}.

This difficulty can be avoided by introducing a dual description of the system in terms of the gaps between the particles instead of their positions~\cite{Rizkallah:2022}. This approach has been used to study a driven tracer in the symmetric simple exclusion process in~\cite{Dandekar:2022,Grabsch:2023b}. We apply here this formalism to general systems.

We introduce the dual density $\rd$ as the typical spacing between the particles around particle $z$ and the associated dual current~\cite{Rizkallah:2022}
\begin{equation}
    \label{eq:DefDualFieldsSM}
    \rd(z,t) = \frac{1}{\rho(x(z,t),t)}
    \:,
    \quad
    \jd(z,t) = - \frac{j(x(z,t),t)}{\rho(x(z,t),t)}
    \:,
\end{equation}
where we have denoted $x(z,t)$ the position of particles $z$ at time $t$. It can be determined as
\begin{equation}
    \label{eq:DefPosSM}
    x(z,t) = \int_0^z \rd(z',t) \dd z'
    - \int_0^t \jd(0,t') \dd t'
    \:.
\end{equation}
The inverse transformation takes the same form,
\begin{equation}
    \rho(x,t) = \frac{1}{\rd(z(x,t),t)}
    \:,
    \quad
    j(x,t) = - \frac{\jd(z(x,t),t)}{\rd(z(x,t),t)}
    \:,
\end{equation}
\begin{equation}
    z(x,t) = \int_0^x \rho(x',t) \dd x'
    - \int_0^t j(0,t') \dd t'
    \:.
\end{equation}
Remarkably, these dual fields obey the same fluctuating hydrodynamics equations~\eqref{eq:fhSM},
\begin{equation}
  \partial_t \rd+\partial_z \jd = 0
  \:,
  \qquad
  \jd=- \tilde{D}(\rd)\partial_z \rd +
  \sqrt{\frac{\tilde\sigma(\rd)}{\Lambda}}\,\eta \:,
  \label{eq:fhDualSM}
\end{equation}
but with dual transport coefficients
\begin{equation}
    \label{eq:DualTrSM}
    \tilde{D}(\rho) = \frac{1}{\rho^2} D \left( \frac{1}{\rho} \right)
    \:,
    \quad
    \tilde{\sigma}(\rho) = \rho \: \sigma \left( \frac{1}{\rho} \right)
    \:.
\end{equation}
In this dual picture, the displacement of the tracer (the particle with label $z=0$) is given by
\begin{equation}
    \label{eq:DisplTrDualSM}
    Y_t = x(0,t) = - \int_0^t \tilde{j}(0,t') \dd t'
    \equiv - \tilde{Q}_t
    \:,
\end{equation}
where $\tilde{Q}_t$ is the integrated current through the origin in the dual system. This is a natural relation, since the displacement of the tracer is the opposite of the ``flux of gaps'' through the tracer.

The key point is that, in this dual picture, the labels of the particles play the role of the spatial coordinate. Hence, the tracer being the particle with label $0$ remains fixed at the origin of this new reference frame.
Consequently, the pressure jump~\eqref{eq:MatchBiasSM} becomes a discontinuity of the dual chemical potential  $\tilde{\mu}$,
\begin{equation}
    \label{eq:DiscMuDualSM}
    \tilde{\mu}\big(
    \rd(0^+,t)
    \big)
    - \tilde{\mu}\big(
    \rd(0^-,t)
    \big)
    = - F
    \:,
    \qquad
    \partial_\rho \tilde\mu(\rho)
    = \frac{2 k_{\mathrm{B}} T \tilde{D}(\rho)}{\tilde\sigma(\rho)}
    \:.
\end{equation}
Additionally, the second matching condition~\eqref{eq:ContXtSM} becomes through the duality transformation~\eqref{eq:DefDualFieldsSM}
\begin{equation}
    \label{eq:ContJDualSM}
    \jd(0^+,t) = \jd(0^-,t)
    \:.
\end{equation}
In this dual picture, the tracer being fixed at the origin, the matching conditions that enforce the driving force $F$ into the macroscopic formalism~\eqref{eq:DiscMuDualSM}-\eqref{eq:ContJDualSM} are now expressed at the origin only. This solves the difficulty of following the tracer. The displacement of the tracer is then deduced from~\eqref{eq:DisplTrDualSM}, which using the continuity relation~\eqref{eq:fhDualSM} can be expressed as
\begin{equation}
    Y_t = - \int_0^\infty \big[
    \rd(z,t) - \rd(z,0)
    \big]
    \dd z
    \:.
\end{equation}
The evolution equations~\eqref{eq:fhDualSM} with the matching conditions~\eqref{eq:DiscMuDualSM}-\eqref{eq:ContJDualSM} thus fully determine the evolution of the position of the tracer.

\section{Field theory and saddle point}

The evolution equations~\eqref{eq:fhDualSM} together with the conditions~\eqref{eq:DiscMuDualSM}-\eqref{eq:ContJDualSM} can be rewritten to express the probability of observing a given evolution of the density from a ninitial density $\rd(z,0)$ to a final density $\rd(z,T)$ at a time $T$. Using a standard Martin-Siggia-Rose-Janssen-De Dominicis-Peliti
approach~\cite{Martin:1973,Janssen:1976,DeDominicis:1978}, this gives
\begin{multline}
    \mathbb{P}[ \{ \rd(z,t) \}_{t \in [0,T]} ]
    \propto
    \int \D \jd \: \exp\left[
    - \Lambda \int_{\mathbb{R}^\star} \dd z \int_0^T \dd t \frac{(\jd + \tilde{D}(\rd) \partial_z \rd)^2}{2 \tilde{\sigma}(\rd)}
    \right]
    \prod_{x,t} \delta \big( \partial_t \rd + \partial_z \jd \big)
    \\
    \times
    \prod_t \delta \big(
    \tilde{\mu}\big(
    \rd(0^+,t)
    \big)
    - \tilde{\mu}\big(
    \rd(0^-,t)
    \big) + F
    \big)
    \prod_t
    \delta \big(
    \jd(0^+,t) - \jd(0^-,t)
    \big)
    \:,
\end{multline}
where we have inserted the expression of the current~\eqref{eq:fhDualSM} into the Gaussian measure of the noise $\eta$, and integrated over the possible evolutions of the current $\jd$. Expressing the delta-functions as integrals, we can rewrite this probability as
\begin{equation}
    \label{eq:ProbEvolSM}
    \mathbb{P}[ \{ \rd(z,t) \}_{t \in [0,T]} ]
    \propto
    \int \D \jd(z,t) \D H(z,t) \D a(t) \D b(t)
    \:
    \e^{- \Lambda S_T[\rd, \jd, H, a, b]}
    \:,
\end{equation}
where we introduced the action
\begin{multline}
    \label{eq:ActionDualSM}
    S_T[\rd, \jd, H, a, b]
    = \int_{\mathbb{R}^\star} \dd z \int_0^T \dd t
    \left[ \frac{(\jd + \tilde{D}(\rd) \partial_z \rd)^2}{2 \tilde{\sigma}(\rd)}
    + H(\partial_t \rd + \partial_z \jd)
    \right]
    \\
    + \int_0^T \dd t \Big[
    a(t) \Big(
    \tilde{\mu}\big(
    \rd(0^+,t)
    \big)
    - \tilde{\mu}\big(
    \rd(0^-,t)
    \big) + F
    \Big)
    + b(t) \Big(
    \jd(0^+,t) - \jd(0^-,t)
    \Big)
    \Big]
    \:.
\end{multline}
This formulation allows us to analyze the distribution of various observables, and in particular the displacement $X_t$ of the tracer.

\subsection{Cumulant generating function}

The action formulation~\eqref{eq:ProbEvolSM} allows us to write the moment generating function of the displacement $X_T$ of the tracer as
\begin{equation}
    \label{eq:MomGenSM}
    \moy{\e^{\lambda X_T}}
    = \moy{\e^{-\Lambda \lambda \tilde{Q}_{T/\Lambda^2}}}
    = \frac{\displaystyle \int \D \rd \D \jd \D H \D a \D b \: \mathbb{P}[\rd(z,0)]
    \: \e^{- \Lambda S_{T/\Lambda^2}[\rd,\jd,H,a,b] - \Lambda \lambda \tilde{Q}_{T/\Lambda^2} }
    }
    {\displaystyle
    \int \D \rd \D \jd \D H \D a \D b \: \mathbb{P}[\rd(z,0)]
    \: \e^{- \Lambda S_{T/\Lambda^2}[\rd,\jd,H,a,b]}
    }
    \:,
\end{equation}
where we have introduced the distribution of the initial density $\mathbb{P}[\rd(z,0)]$. The precise form of this distribution depends on the choice under consideration. The two main classes are typically
\begin{itemize}
    \item ``annealed'' initial conditions in which the system is locally at equilibrium at a density $\rd_{\mathrm{i}}(z)$, for which~\cite{Derrida:2009a,Derrida:2025a}
    \begin{equation}
        \mathbb{P}[\rd(z,0)]
        \propto
        \exp \left[- \Lambda \int \dd z
        \int_{\rd_{\mathrm{i}}}^{\rd(z,0)} \dd r
        (\rho(z,0) - r) 
        \frac{2 \tilde{D}(r)}{\tilde{\sigma}(r)}
        \right]
        \:,
    \end{equation}
    \item ``quenched'' initial conditions in which the initial density is fixed
    \begin{equation}
        \mathbb{P}[\rd(z,0)]
        = \prod_x \delta \big( \rd(z,0) - \rd_{\mathrm{i}}(z) \big)
        \:.
    \end{equation}
\end{itemize}
In the following, since we will only focus on the mean displacement, the choice of the initial condition will be irrelevant.

From the expression of the moment generating function, it is natural to make the choice $\Lambda = \sqrt{T} \gg 1$. The integrals being dominated by the minimum of the action, we get from a saddle point estimate
\begin{equation}
    \label{eq:CumulSM}
    \ln \moy{\e^{\lambda X_T}}
    \underset{T \to \infty}{\simeq}
    -\sqrt{T} \left[
    S_1[q,k,p,\alpha,\beta] + \lambda \int_{0}^\infty [q(z,1) - q(z,0)] \dd z 
    - S_1[q,k,p,\alpha,\beta] \Big|_{\lambda = 0}
    \right]
    \:,
\end{equation}
where we have denoted $(q,k,p,\alpha,\beta)$ the optimal value of $(\rd,\jd,H,a,b)$ which minimize the action in the numerator of~\eqref{eq:MomGenSM}. The equations satisfied by this saddle point can be derived by looking at small variations around the optimal fields, as done for instance in~\cite{Derrida:2009a,Krapivsky:2014,Krapivsky:2015a,Dandekar:2022,Grabsch:2023b}. As an illustration, we consider the example of the field $\jd$. We evaluate the action at $\jd = k + \delta \jd$ and collect the linear terms in $\delta \jd$. This gives
\begin{equation}
    \int_{\mathbb{R}^\star} \dd z \int_0^1 \dd t \left[ \delta \jd
    \frac{k + \tilde{D}(q) \partial_z q}{\tilde\sigma(q)}
    + p \partial_z \delta \jd
    \right]
    + \int_0^1 \dd t \: \beta(t) \big(
    \delta \jd(0^+,t) - \delta \jd(0^-,t)
    \big)
    \:.
\end{equation}
Performing an integration by parts in the first term and imposing that the terms in $\delta \jd$ vanish, we obtain
\begin{equation}
    \label{eq:SaddlePtJSM}
    k = - \tilde{D}(q) \partial_z q + \tilde\sigma(q) \partial_z p
    \:,
    \qquad
    p(0^+,t) = p(0^-,t) = \beta(t)
    \:.
\end{equation}
Proceeding similarly for the other fields gives 
\begin{equation}
    \partial_t q + \partial_z k = 0
    \:,
    \qquad
    \partial_t p = - \tilde{D}(q) \partial_z^2 p
    - \frac{\tilde\sigma'(q)}{2}(\partial_z p)^2
    \:,
    \qquad
    a(t) = \frac{\tilde\sigma(q) \partial_z p}{2} \Big|_{0^+} = \frac{\tilde\sigma(q) \partial_z p}{2} \Big|_{0^-}
    \:,
\end{equation}
\begin{equation}
    \tilde{\mu}\big( q(0^+,t) \big)
    - \tilde{\mu}\big( q(0^-,t) \big)
    = - F
    \:,
    \quad
    k(0^+,t) = k(0^-,t)
    \:,
    \quad
    p(z,1) = - \lambda \Theta(z)
    \:,
\end{equation}
and an initial condition that depends on the choice of the initial distribution
\begin{equation}
    \label{eq:InitCondSM}
    \left\lbrace
    \begin{array}{lll}
        p(z,0) &= - \lambda \Theta(z) +
        \frac{1}{k_{\mathrm{B}}T} \left[ \tilde\mu \big( q(z,0) \big) - \tilde\mu \big( \tilde{\rho}_{\mathrm{i}}(z)  \big)
        \right] &
        \text{(annealed)}\\
        q(z,0) &= \tilde{\rho}_{\mathrm{i}}(z) 
        & \text{(quenched)}
    \end{array}
    \right.
\end{equation}
Combining these equations, we get the usual equations of macroscopic fluctuation theory~\cite{Derrida:2009a}
\begin{equation}
    \label{eq:MFTSM}
    \partial_t q = \partial_z [\tilde{D}(q) \partial_z q - \tilde\sigma(q) \partial_z p]
    \:,
    \qquad
    \partial_t p = - \tilde{D}(q) \partial_z^2 p
    - \frac{\tilde\sigma'(q)}{2}(\partial_z p)^2
    \:,
\end{equation}
with the initial condition~\eqref{eq:InitCondSM} and the final condition $p(z,1) = - \lambda \Theta(z)$, but supplemented by the boundary conditions at the origin
\begin{equation}
    \label{eq:MatchMFTSM}
    \tilde{\mu}\big( q(0^+,t) \big)
    - \tilde{\mu}\big( q(0^-,t) \big)
    = - F
    \:,
    \quad
    p(0^+,t) = p(0^-,t)
    \:,
    \quad
    \big[ \tilde{D}(q) \partial_z q \big]_{0^-}^{0^+} = 0
    \:,
    \quad
    \big[ \tilde{\sigma}(q) \partial_z p \big]_{0^-}^{0^+} = 0
    \:.
\end{equation}
These equations fully determine the saddle point fields $(q,p)$. In particular, since the cumulant generating function~\eqref{eq:CumulSM} is obtained by a minimization over all the fields, its derivative takes a compact form
\begin{equation}
    \label{eq:DerCumulSM}
    \dt{}{\lambda} 
    \ln \moy{\e^{\lambda X_T}}
    \underset{T \to \infty}{\simeq}
    - \sqrt{T} \int_0^\infty \big[
    q(z,1) - q(z,0)
    \big] \dd z
    \:,
\end{equation}
from which the cumulants can be computed.

\subsection{Mean displacement}

The mean displacement of the tracer can be deduced from the solution of the macroscopic fluctuation theory equations~\eqref{eq:MFTSM}-\eqref{eq:DerCumulSM} by evaluating them at $\lambda = 0$. This gives
\begin{equation}
    \moy{X_T} \underset{T \to \infty}{\simeq} 
    - \sqrt{T}
    \int_0^\infty \big[
    q_0(z,1) - q_0(z,0)
    \big] \dd z
    \:,
\end{equation}
where we have denoted $q_0 = q |_{\lambda = 0}$, which obeys
\begin{equation}
    \label{eq:EqsForMeanXtSM}
    \partial_t q_0 = \partial_z [\tilde{D}(q_0) \partial_z q_0]
    \:,
    \quad
    \tilde{\mu}\big( q_0(0^+,t) \big)
    - \tilde{\mu}\big( q_0(0^-,t) \big)
    = - F
    \:,
    \quad
     \big[ \tilde{D}(q_0) \partial_z q_0 \big]_{0^-}^{0^+} = 0
     \:,
     \quad 
    q_0(z,0) = \tilde{\rho}_{\mathrm{i}}(z)
    \:.
\end{equation}
If the initial density is a step function
\begin{equation}
    \tilde{\rho}_{\mathrm{i}}(z) = 
    \frac{1}{\rho_+} \Theta(z)
    + \frac{1}{\rho_-} \Theta(-z)
    \:,
\end{equation} 
we can look for a self-similar solution of these equations,
\begin{equation}
    q_0(z,t) = 
    Q \left( u = \frac{z}{\sqrt{t}} \right)
    \:,
\end{equation}
which reduce to
\begin{equation}
    \label{eq:QSelfSimilSM}
    \dt{}{u} \left[ \tilde{D}(Q) \dt{Q}{u} \right]
    + \frac{u}{2} \dt{Q}{u}
    = 0
    \:,
    \quad
    \big[ \tilde{\mu}\big( Q \big) \big]_{0^-}^{0^+}
    = - F
    \:,
    \quad
    \big[ \tilde{D}(Q) Q' \big]_{0^-}^{0^+}
    = 0
    \:,
    \quad
    Q(\pm \infty) = \frac{1}{\rho_\pm}
    \:.
\end{equation}
The mean displacement is then
\begin{equation}
    \label{eq:MeanDisplXtFinalSM}
    \lim_{T \to \infty} \frac{\moy{X_T}}{\sqrt{T}}
    = - \int_{0}^\infty \left[ Q(u) - \frac{1}{\rho_+} \right] \dd u
    = 2 \tilde{D}(Q) Q' \Big|_{0}
    \:,
\end{equation}
where we have used the first equation in~\eqref{eq:QSelfSimilSM} to simplify the integral.
These are the equations given in the main text.

\subsection{Mean density profile}

The mean density profile can be computed from the duality mapping~\eqref{eq:DefDualFieldsSM}. Indeed, since this mapping holds for any realization of the fields, it holds in particular for the typical realization which is the saddle point. Hence, we get
\begin{equation}
    \label{eq:MeanDensSM}
    \moy{\rho(X_t + x_0(z,t), t)} = \frac{1}{q_0(z,t)}
    \:,
    \quad
    x_0(z,t) = \int_0^z q_0(z',t) \dd z'
    \:.
\end{equation}
These are the relations given in the main text.

\subsection{Beyond the mean: the example of a specific model}

Solving the MFT equations~\eqref{eq:MFTSM} with the matching conditions~\eqref{eq:MatchMFTSM} to get the cumulant generating function~\eqref{eq:DerCumulSM} is a challenging task, even for minimal models like the SEP, corresponding to $D(\rho) = 1$ and $\sigma(\rho) = 2 \rho(1-\rho)$.

It is nevertheless possible to obtain an explicit solution of these equations for specific choices of $D$ and $\sigma$. For instance, consider
\begin{equation}
    \label{eq:SpecChoiceTrCoefs}
    D(\rho) = \frac{D_0}{\rho^2}
    \:,
    \quad
    \sigma(\rho) = a + b \rho
    \:,
\end{equation}
which maps via the duality transform~\eqref{eq:DualTrSM} onto
\begin{equation}
    \label{eq:TrDualSolvable}
    \tilde{D}(\rho) = D_0
    \:,
    \quad
    \tilde{\sigma}(\rho) = a \rho + b
    \:.
\end{equation}
For $b = 0$ and $a = 2D_0$, the dual model corresponds to reflecting Brownian particles~\cite{Krapivsky:2014,Krapivsky:2015a}. In this case, Eqs.~\eqref{eq:MFTSM}-\eqref{eq:DerCumulSM} thus describe the cumulants of the current through the origin in the presence of a bias at the origin.

For the specific choice of transport coefficients~\eqref{eq:TrDualSolvable}, the MFT equations~\eqref{eq:MFTSM} become
\begin{equation}
    \label{eq:MFTSMspecific}
    \partial_t q = \partial_z [D_0 \partial_z q - (a q + b) \partial_z p]
    \:,
    \qquad
    \partial_t p = - D_0 \partial_z^2 p
    - \frac{a}{2}(\partial_z p)^2
    \:,
\end{equation}
with the boundary conditions
\begin{equation}
    \label{eq:MatchMFTSMspecific}
    \frac{2 k_{\mathrm{B}} T D_0}{a}
    \ln \frac{a q(0^+,t) + b}{a q(0^-,t) + b}
    = - F
    \:,
    \quad
    p(0^+,t) = p(0^-,t)
    \:,
    \quad
    \big[ \partial_z q \big]_{0^-}^{0^+} = 0
    \:,
    \quad
    \big[ (a q + b) \partial_z p \big]_{0^-}^{0^+} = 0
    \:,
\end{equation}
and initial/final conditions~\eqref{eq:InitCondSM},
\begin{equation}
    p(z,1) = - \lambda \Theta(z)
    \:,
    \quad
    p(z,0) = - \lambda \Theta(z)
    + \frac{2 D_0}{a}
    \ln \frac{a q(z,0) + b}{a \tilde\rho_{\mathrm{i}}(z) + b}
    \:,
    \quad
    \tilde\rho_{\mathrm{i}}(z) = \frac{1}{\rho_-} \Theta(-z) + \frac{1}{\rho_+} \Theta(z)
    \:.
\end{equation}
in the annealed case.

The equations can be simplified using a Cole-Hopf transformation~\cite{Krapivsky:2014,Krapivsky:2015a}
\begin{equation}
    P = \e^{ \frac{a}{2 D_0} p}
    \:,
    \quad
    Q = (a q + b) \e^{ -\frac{a}{2 D_0} p}
    \:,
\end{equation}
which maps them onto
\begin{equation}
    \partial_t Q = D_0 \partial_z^2 Q
    \:,
    \quad
    \partial_t P = - D_0 \partial_z^2 P
    \:,
    \quad
    Q(z,0) = (a \tilde\rho_{\mathrm{i}}(z) + b) \e^{\frac{a \lambda}{2 D_0} \Theta(z)}
    \:,
    \quad
    P(z,1) = \e^{-\frac{a \lambda}{2 D_0} \Theta(z)}
    \:,
\end{equation}
with the boundary conditions
\begin{equation}
    P(0^+,t) = P(0^-,t)
    \:,
    \quad
    Q(0^+,t) = Q(0^-,t) \e^{- \frac{a}{2D_0} \frac{F}{k_{\mathrm{B}T}}}
    \:,
    \quad
    \left[ \partial_z Q \right]_{0^-}^{0^+} = 0
    \:,
    \quad
    \partial_z P(0^+,t)
    = \e^{\frac{a}{2D_0} \frac{F}{k_{\mathrm{B}T}}} \partial_z P(0^-,t)
    \:.
\end{equation}
The equations on $P$ and $Q$ are decoupled, so they can be solved independently.
The solutions take the form
\begin{equation}
    P(z,t) = 1 
    + \frac{1}{2}(\e^{-\frac{a \lambda}{2 D_0}} - 1)
    \erfc \left(
    - \frac{z}{\sqrt{4D_0(1-t)}}
    \right)
    - \frac{1}{2} (\e^{-\frac{a \lambda}{2 D_0}} - 1)
    \tanh \left( \frac{a F}{4 D_0 k_{\mathrm{B}} T} \right)
    \erfc \left(
    \frac{|z|}{\sqrt{4D_0(1-t)}}
    \right)
    \:,
\end{equation}
\begin{multline}
    Q(z,t) = \e^{\frac{a\lambda}{2 D_0}}
    \left( \frac{a}{\rho_+} + b \right) + \frac{1}{2} \left[ 
    \frac{a}{\rho_-} + b - \e^{\frac{a\lambda}{2 D_0}} \left( \frac{a}{\rho_+} + b \right) 
    \right]
    \erfc \left(
        \frac{z}{\sqrt{4 D_0 t}}
    \right)
    \\
    - \frac{\frac{a}{\rho_-} + b + \e^{\frac{a\lambda}{2 D_0}}
    \left( \frac{a}{\rho_+} + b \right) }{2} \tanh \left( \frac{a F}{4 D_0 k_{\mathrm{B}} T} \right)
    \sg{z} 
    \erfc \left(
    \frac{|z|}{\sqrt{4D_0(1-t)}}
    \right)
    \:.
\end{multline}
The cumulant generating function of $X_T$ is then deduced from~\eqref{eq:DerCumulSM}, which becomes
\begin{equation}
    \hat\psi(\lambda) \equiv
    \lim_{T \to \infty}
    \frac{1}{\sqrt{T}} \ln \moy{\e^{\lambda X_T}}
    \:,
    \quad
    \dt{\hat\psi}{\lambda}
    = - \frac{1}{a} \int_0^\infty \left[
    Q(z,1)P(z,1) - Q(z,0)P(z,0)
    \right] \dd z
    \:.
\end{equation}
This gives
\begin{equation}
    \hat\psi(\lambda)
    = \frac{4 D_0}{a^2} \sqrt{\frac{D_0}{\pi}}
    \left[
     \left( \frac{a}{\rho_+} + b \right)\frac{\e^{\frac{a\lambda}{2 D_0}}-1}{1 + \e^{-\frac{a F}{2 D_0 k_{\mathrm{B}} T}}}
    +  \left( \frac{a}{\rho_-} + b \right)\frac{\e^{-\frac{a\lambda}{2 D_0}}-1}{1 + \e^{\frac{a F}{2 D_0 k_{\mathrm{B}} T}}}
    \right]
    \:.
\end{equation}
This result, obtained for the specific choice of transport coefficients~\eqref{eq:SpecChoiceTrCoefs}, shows that the formalism presented in this article gives access to the full distribution of the displacement of the tracer, upon solving the MFT equations~\eqref{eq:MFTSM}-\eqref{eq:DerCumulSM}.

\section{Numerical resolution}

The equations for the mean displacement of the tracer~\eqref{eq:QSelfSimilSM} cannot be solved analytically for generic $D$ and $\sigma$, so we must resort to a numerical calculation. We sketch here the procedure we used to compute the solution of the equations for the scaling function $Q$~\eqref{eq:QSelfSimilSM}.

The problem is parameterized by $\rho_+$, $\rho_-$ and $F$, and one must determine $Q(0^\pm)$ and $Q'(0^\pm)$ to compute the mean displacement~\eqref{eq:MeanDisplXtFinalSM}. It is actually simpler to use $\rho_+$, $\rho_-$ and $Q(0^+)$ as parameters and compute $Q(0^-)$, $Q'(0^\pm)$ and $F$. Our algorithm is as follows.
\begin{enumerate}
    \item For a given choice of $\rho_+$ and $\rho_-$, choose a value of $Q(0^+)$.
    \item Solve the differential equation~\eqref{eq:QSelfSimilSM} on $[0,L]$ (for $L$ large enough, typically $L=25$), with initial value $Q(0^+)$ and a guess on $Q'(0^+)$. Adjust the guess until the solution satisfies $Q(L) = 1/\rho_+$. This gives the value of $Q'(0^+)$.
    \item Select a guess for $Q(0^-)$ and deduce $Q'(0^-)$ from the relation of the derivatives~\eqref{eq:QSelfSimilSM}. Solve the differential equation~\eqref{eq:QSelfSimilSM} on $[-L,0]$ and adjust the guess on $Q(0^-)$ until $Q(-L) = 1/\rho_-$. This gives both $Q(0^-)$ and $Q'(0^-)$.
    \item Determine the force $F$ from the discontinuity at the origin~\eqref{eq:QSelfSimilSM}.
\end{enumerate}
We applied this procedure for different values of $Q(0^+)$ to obtain the plots of the mean displacement as a function of $F$ shown in the main text.

\section{Small force expansion}

For generic systems, one can get exact results in the small force limit, when the initial density is flat with $\rho_+ = \rho_- = \bar\rho$. Denoting $\bar{\tilde{\rho}} = 1/\bar\rho$, we can look for a solution of the equations~\eqref{eq:QSelfSimilSM} in powers of the force $F$,
\begin{equation}
    Q(u) = \bar{\tilde{\rho}}
    + \sum_{n=1}^\infty \left( \frac{F}{k_{\mathrm{B}}T} \right)^n Q_n(u)
    \:.
\end{equation}
Note that due to the symmetry of the problem under $u \to -u$ and $F \to -F$, we have that $Q_n(-u) = (-1)^n Q_n(u)$. We can thus focus on the solution for $u>0$. 

At first order, the solution takes the form
\begin{equation}
    Q_1(u)
    = - \frac{\tilde{\sigma}(\bar{\tilde{\rho}})}{4 \tilde{D}(\bar{\tilde{\rho}})}
    \erfc \left(
    \frac{u}{\sqrt{4 \tilde{D}(\bar{\tilde{\rho}})}}
    \right)
    \:.
\end{equation}
In terms of the original transport coefficients~\eqref{eq:DualTrSM} this gives
\begin{equation}
    Q_1(u) = - \frac{\sigma(\bar{\rho})}{4 \bar\rho^3 D(\bar\rho)}
    \erfc \left(
    \frac{u}{\sqrt{4 \bar\rho^2 D(\bar\rho)}}
    \right)
\end{equation}
Combined with the expression of the mean displacement~\eqref{eq:MeanDisplXtFinalSM}, this gives 
\begin{equation}
    \lim_{T \to \infty} \frac{\moy{X_T}}{\sqrt{T}}
    = \frac{\sigma(\bar\rho)}{\bar\rho^2 \sqrt{4 \pi D(\bar\rho)}} \frac{F}{k_{\mathrm{B}}T} + O(F^2)
    \:,
\end{equation}
as announced in the main text.

Pushing the expansion further in $F$, we get
\begin{multline}
    Q_2(u) = \frac{\sigma(\bar\rho)^2}{32 \bar\rho^5 D(\bar\rho)^3}
    \Bigg[
    \erf \left(
    \frac{u}{\sqrt{4 \bar\rho^2 D(\bar\rho)}}
    \right)^2
    - \erf \left(
    \frac{u}{\sqrt{4 \bar\rho^2 D(\bar\rho)}}
    \right)
    + \frac{2}{\pi} \e^{- \frac{u^2}{2\bar\rho^2 D(\bar\rho)}}
    \\
    - \frac{u}{\bar\rho \sqrt{\pi D(\bar\rho)}}
    \e^{- \frac{u^2}{4 \bar\rho^2 D(\bar\rho)}}
    \erfc \left(
    \frac{u}{\sqrt{4 \bar\rho^2 D(\bar\rho)}}
    \right)
    \Bigg]
    \:,
\end{multline}
\begin{multline}
    Q_3(u) = 
    \frac{e^{-2 U^2} \left(12 \sqrt{\pi } \bar{\rho }^5 \sigma
   \left(\bar{\rho }\right)^3 D'\left(\bar{\rho }\right)^2+48
   \sqrt{\pi } \bar{\rho }^4 D\left(\bar{\rho }\right) \sigma
   \left(\bar{\rho }\right)^3 D'\left(\bar{\rho }\right)+48 \sqrt{\pi
   } \bar{\rho }^3 D\left(\bar{\rho }\right)^2 \sigma \left(\bar{\rho
   }\right)^3\right)}{768 \pi ^{3/2} \bar{\rho }^{10}
   D\left(\bar{\rho }\right)^5}
    \\
   + \frac{e^{-3 U^2} \left(-6 U \bar{\rho }^5 \sigma \left(\bar{\rho
   }\right)^3 D'\left(\bar{\rho }\right)^2-24 U \bar{\rho }^4
   D\left(\bar{\rho }\right) \sigma \left(\bar{\rho }\right)^3
   D'\left(\bar{\rho }\right)-24 U \bar{\rho }^3 D\left(\bar{\rho
   }\right)^2 \sigma \left(\bar{\rho }\right)^3\right)}{768 \pi
   ^{3/2} \bar{\rho }^{10} D\left(\bar{\rho }\right)^5}
   \\
   +
   \frac{\text{erfc}(U)^3 \left(2 \pi ^{3/2} \bar{\rho }^5
   D\left(\bar{\rho }\right) \sigma \left(\bar{\rho }\right)^3
   D''\left(\bar{\rho }\right)+12 \pi ^{3/2} \bar{\rho }^4
   D\left(\bar{\rho }\right) \sigma \left(\bar{\rho }\right)^3
   D'\left(\bar{\rho }\right)+12 \pi ^{3/2} \bar{\rho }^3
   D\left(\bar{\rho }\right)^2 \sigma \left(\bar{\rho
   }\right)^3\right)}{768 \pi ^{3/2} \bar{\rho }^{10}
   D\left(\bar{\rho }\right)^5}
   \\
   +
   \frac{\text{erfc}\left(\sqrt{3} U\right) \left(9 \sqrt{3 \pi }
   \bar{\rho }^5 \sigma \left(\bar{\rho }\right)^3 D'\left(\bar{\rho
   }\right)^2-6 \sqrt{3 \pi } \bar{\rho }^5 D\left(\bar{\rho }\right)
   \sigma \left(\bar{\rho }\right)^3 D''\left(\bar{\rho
   }\right)\right)}{768 \pi ^{3/2} \bar{\rho }^{10} D\left(\bar{\rho}\right)^5}
   +
   \\
   \text{erfc}(U)^2 \Bigg[
   \text{erf}(U)
   \frac{6 \pi ^{3/2} \bar{\rho}^5 \sigma \left(\bar{\rho }\right)^3 D'\left(\bar{\rho }\right)^2
   +24\pi ^{3/2} \bar{\rho}^4 D\left(\bar{\rho }\right)
   \sigma \left(\bar{\rho }\right)^3 D'\left(\bar{\rho}\right)+ 24\pi ^{3/2} \bar{\rho }^3  D\left(\bar{\rho }\right)^2 \sigma \left(\bar{\rho }\right)^3}{768 \pi ^{3/2} \bar{\rho}^{10}   D\left(\bar{\rho }\right)^5}
   \\
   - \frac{e^{-U^2} U \sigma \left(\bar{\rho }\right)^3 \left(\left(2
   U^2-9\right) \bar{\rho }^2 D'\left(\bar{\rho }\right)^2+2
   \bar{\rho } D\left(\bar{\rho }\right) \left(\bar{\rho }
   D''\left(\bar{\rho }\right)+4 \left(U^2-3\right) D'\left(\bar{\rho
   }\right)\right)+8 \left(U^2-3\right) D\left(\bar{\rho
   }\right)^2\right)}{256 \sqrt{\pi } \bar{\rho }^7 D\left(\bar{\rho}\right)^5}
   \Bigg]
   \\
   + \erfc(U) \Bigg[
   \frac{e^{-2 U^2} \sigma \left(\bar{\rho }\right)^3
   \left(\left(U^2-3\right) \bar{\rho }^2 D'\left(\bar{\rho
   }\right)^2+\bar{\rho } D\left(\bar{\rho }\right) \left(\bar{\rho }
   D''\left(\bar{\rho }\right)+\left(4 U^2-6\right) D'\left(\bar{\rho
   }\right)\right)+\left(4 U^2-6\right) D\left(\bar{\rho
   }\right)^2\right)}{64 \pi  \bar{\rho }^7 D\left(\bar{\rho}\right)^5}
    \\
   -\frac{e^{-U^2} U \sigma \left(\bar{\rho }\right)^3 \left(\bar{\rho }
   D'\left(\bar{\rho }\right)+2 D\left(\bar{\rho
   }\right)\right)^2}{64 \sqrt{\pi } \bar{\rho }^7 D\left(\bar{\rho}\right)^5}
   +
   \frac{\sigma \left(\bar{\rho }\right)}{768 \pi  \bar{\rho }^7 D\left(\bar{\rho
   }\right)^5} \Big(
   3 \left(4-3\sqrt{3}\right) \bar{\rho }^2 \sigma\left(\bar{\rho }\right)^2
   D'\left(\bar{\rho }\right)^2
   \\
   +2 \bar{\rho} D\left(\bar{\rho}\right) \sigma \left(\bar{\rho}\right) \left(\sigma
   \left(\bar{\rho }\right) \left(3 \left(\sqrt{3}-2\right) \bar{\rho
   } D''\left(\bar{\rho }\right)+ 2 (\pi -9)D'\left(\bar{\rho}\right)\right)-2 (\pi -3) \bar{\rho } D'\left(\bar{\rho}\right)
   \sigma '\left(\bar{\rho }\right)\right)
   \\
   +2 D\left(\bar{\rho
   }\right)^2 \left(2 \pi  \bar{\rho }^2 \sigma '\left(\bar{\rho
   }\right)^2-\bar{\rho } \sigma \left(\bar{\rho }\right) \left(\pi 
   \bar{\rho } \sigma ''\left(\bar{\rho }\right)+4 (2 \pi -3) \sigma
   '\left(\bar{\rho }\right)\right)+6 (\pi -4) \sigma \left(\bar{\rho
   }\right)^2\right)\Big)
   \Bigg]
    \:.
\end{multline}
From these expressions we deduce using~\eqref{eq:MeanDisplXtFinalSM}
\begin{multline}
    \lim_{T \to \infty} \frac{\moy{X_T}}{\sqrt{T}}
    = \frac{\sigma(\bar\rho)}{\bar\rho^2 \sqrt{4 \pi D(\bar\rho)}} \frac{F}{k_{\mathrm{B}}T}
    + \Bigg[
    \frac{\sigma (\bar\rho )^3 \left(D'(\bar\rho )^2-2 D(\bar\rho ) D''(\bar\rho)\right)}{256 \sqrt{\pi } \bar\rho ^4 D(\bar\rho )^{9/2}}
    + \frac{\sqrt{3} \sigma (\bar\rho )^3 \left(3 D'(\bar\rho )^2-2 D(\bar\rho )
   D''(\bar\rho )\right)}{128 \pi ^{3/2} \bar\rho ^4 D(\bar\rho )^{9/2}}
   \\
   + \frac{3 \sigma (\bar\rho )^3 \left(D(\bar\rho ) D''(\bar\rho )-D'(\bar\rho)^2\right)}{64 \pi ^{3/2} \bar\rho ^4 D(\bar\rho )^{9/2}}
   +
   \frac{(\pi -3) \sigma (\bar\rho )^2 D'(\bar\rho ) \left(\bar\rho  \sigma '(\bar\rho)- 4 \sigma (\bar\rho )\right)}{96 \pi ^{3/2} \bar\rho ^5 D(\bar\rho )^{7/2}}
   + \frac{\sigma (\bar\rho )^2 \sigma ''(\bar\rho )}{192 \sqrt{\pi } \bar\rho ^4
   D(\bar\rho )^{5/2}}
   \\
   -\frac{\sigma (\bar\rho ) \sigma '(\bar\rho )^2}{96
   \sqrt{\pi } \bar\rho ^4 D(\bar\rho )^{5/2}}
    +
    \frac{(5-2 \pi ) \sigma (\bar\rho )^3}{32 \pi ^{3/2} \bar\rho ^6 D(\bar\rho)^{5/2}}
    +\frac{(2 \pi -3) \sigma (\bar\rho )^2 \sigma'(\bar\rho )}{48 \pi^{3/2} \bar\rho^5 D(\bar\rho )^{5/2}}
    \Bigg] \left( \frac{F}{k_{\mathrm{B}}T} \right)^3 + 
    O(F^5)
    \:,
\end{multline}
since the even orders vanish by symmetry.

\section{Numerical simulations}

\subsection{Brownian particles in one dimension}

The simulations of interacting Brownian particles in one dimension are performed by a direct discretization of the evolution equations
\begin{equation}
  \frac{\dd x_n}{\dd t}
  =-\mu_0\sum_{m\neq n}V'(x_n-x_m)+
  \mu_0 F \delta_{n,0}
  +\sqrt{2\mu_0 k_{\mathrm{B}}T}\,\eta_n .
  \label{eq:brownian_microSM}
\end{equation}
The simulations are performed with $N=400$ particles on periodic systems of different lengths $L$ to adjust the mean density $\bar\rho$. Initially the particles are equally spaced. The time discretization is $\delta t = 0.0002$ and the simulations are averaged over $100$ realisations. The other parameters are set to $\mu_0 = k_{\mathrm{B}} = T = 1$. The mean density of particles in the reference frame of the tracer is computed by averaging over $1000$ simulations.

We use the Calogero interaction potential $V(x) = \frac{1}{x^2}$, for which the equation of state and thus the diffusion coefficient can be computed analytically~\cite{Choquard:2000,Lewin:2022,Grabsch:2025b,Grabsch:2026}.

\subsection{Overdamped particles in a quasi-1D channel}

The simulations of particles in a quasi-1D channel are performed with LAMMPS~\cite{Thompson:2022} using $N = 800$ particles in $2d$ channels of different length $L = N/\bar\rho$ and width $1.86$ with a Weeks-Chandler-Andersen potential~\cite{Weeks:1971} which is a smooth approximation of hard spheres. The equations of motion are
\begin{equation}
    m \frac{\dd^2 x_n}{\dd t^2}
    = - \gamma \dt{x_n}{t}
    - \sum_{m \neq n} V'(x_n - x_m)
    + F \delta_{n,0}
    + \sqrt{2 \gamma k_{\mathrm{B}} T} \eta_n
    \:.
\end{equation}
We chose a damping factor $\gamma = 1/\mu_0 = 1$, temperature $T = 1$ and mass $m=0.1$. The measures are averaged over $50$ realizations for the mean displacement and over $1000$ realizations for the density profile. The diffusion coefficient is computed by relying on the transfer matrix method described in~\cite{Benichou:2026}, which assumes nearest-neighbor interaction only. The analytical results are therefore exact for hard disks, but for the WCA potential used in the simulations this becomes an approximation valid at low density.


%